\documentclass[5p]{elsarticle}     %Elsevier format. Also uncomment next line. 
\newproof{IEEEproof}{Proof}
\usepackage[utf8]{inputenc}
\usepackage[english]{babel}
\usepackage[normalem]{ulem}
\usepackage{amsmath}
\usepackage{amssymb}
\usepackage{amsfonts}
\usepackage{fouriernc}
\usepackage{algorithm}
\usepackage{algorithmic}
\usepackage{graphicx}
\usepackage{subfig}
\usepackage{float}
\usepackage{xcolor}
\usepackage{epstopdf}

\newcommand{\bb}[1]{\mathbb{#1}}
\newcommand{\ca}[1]{\mathcal{#1}}

\usepackage{hyperref}
\newtheorem{thm}{Theorem}[section]

\newtheorem{df}[thm]{Definition}
\newtheorem{prob}[thm]{Problem}
\newtheorem{rem}[thm]{Remark}
\newtheorem{pr}[thm]{Proposition}
\newtheorem{eg}[thm]{Example}
\newtheorem{ass}[thm]{Assumption}

\makeatletter
\def\blfootnote{\xdef\@thefnmark{}\@footnotetext}
\makeatother

\usepackage{amssymb}
\begin{document}
\begin{frontmatter}
\title{\Large Resilience to Denial-of-Service and Integrity Attacks:\\ A Structured Systems Approach \tnoteref{t1}%\tnoteref{t1,t2}}
\tnotetext[t1]{Work supported in part by the NSF under Grants $CNS-1446665$ and $CMMI-1362303$, and by the AFOSR under Grant $FA9550-15-10050$. }%Part of this work was carried out in Summer $2015$ when the first author was visiting TCS Innovation Labs, Bangalore.}
%\tnotetext[t2]{A preliminary version of this work was presented at the IEEE American Control Conference, $2016$ \cite{bhaskar2016structres}.
}
\author[eceuw]{Bhaskar Ramasubramanian}
\ead{bhaskarr@uw.edu}
\author[tcs]{M. A. Rajan}
\ead{rajan.ma@tcs.com}
\author[tcs]{M. Girish Chandra}
\ead{m.gchandra@tcs.com}
\author[isr,cs]{Rance Cleaveland}
\ead{rance@cs.umd.edu}
\author[isr,ece]{Steven I. ~Marcus}
\ead{marcus@umd.edu}
\address[eceuw]{Network Security Lab, Department of Electrical and Computer Engineering, University of Washington, Seattle, WA 98195, USA}
\address[tcs]{Innovation Labs, Tata Consultancy Services, Bangalore 560066, Karnataka, India}
\address[isr]{Institute for Systems Research, University of Maryland, College Park, MD 20742, USA}
\address[cs]{Department of Computer Science, University of Maryland, College Park, MD 20742, USA}
\address[ece]{Department of Electrical and Computer Engineering, University of Maryland, College Park, MD 20742, USA}
\begin{abstract} 

The resilience of cyberphysical systems to denial-of-service (DoS)  and integrity attacks is studied in this paper. 
The cyberphysical system is modeled as a linear structured system, and its resilience to an attack is interpreted in a graph theoretical framework. 
The \emph{structural resilience} of the system is characterized in terms of unmatched vertices in maximum matchings of the bipartite graph and connected components of directed graph representations of the system under attack. 
We first present conditions for the system to be resilient to DoS attacks when an adversary may block access or turn off certain inputs to the system. 
We extend this analysis to characterize resilience of the system when an adversary might additionally have the ability to affect the implementation of state-feedback control strategies. 
This is termed an integrity attack. 
We establish conditions under which a system that is structurally resilient to a DoS attack will also be resilient to a certain class of integrity attacks. 
%We show that if a system is structurally resilient to a DoS attack for some defender input matrix $[B_{def}]$ with zero structure $\ca{Z}(B_{def})$, then there exists a $[B'_{def}]$ with $\ca{Z}(B'_{def}) \subseteq \ca{Z}(B_{def})$ for which it will also be structurally resilient to a state feedback integrity attack. 
%If an additional condition holds, we show that the same $[B_{def}]$ will ensure structural resilience to a state feedback integrity attack. 
%
Finally, we formulate an extension to the case of switched linear systems, and derive conditions for such systems to be structurally resilient to a DoS attack. 
\end{abstract}
\end{frontmatter}

\section{Introduction}

Cyberphysical systems (CPSs) are entities in which the working of the physical system is intimately linked to the functioning of computers controlling interactions between the system and a controller, or among subsystems. 
%Since these systems are often controlled via a network, computational resources and bandwidth can also affect their operation. 
%Large scale CPS are ubiquitous in today's world. 
Examples of CPSs include power systems, water distribution networks, medical devices, and automotive systems \cite{baheti2011cyber}. 

Although computer-controlled systems allow for better integration of sensors, actuators, and algorithms, the integrated system is potentially vulnerable to cyber-attacks. 
An attack could be carried out on the physical system, on the computer controlling the system, or on the communication links between the system and the computer. 
The potential scope of attacks on CPSs can be gleaned by an experiment reported in \cite{shoukry2013non}. %, where the Antilock Braking System (ABS) of a vehicle was compromised. 
A spoofer injected a spurious magnetic field that tampered with measurements of speed sensors located on the wheels of the vehicle. 
As a result, the antilock braking system did not work as intended because of the incorrect speed reported to it. 
This attack was completely noninvasive, in that it did not require tampering with sensors on the original system. 
A compilation of vulnerabilities in existing systems, and means of mitigating threats is found in \cite{slay2008lessons}, \cite{farwell2011stuxnet}, \cite{cardenas2008research}. 

\textbf{\emph{Motivation}}: This paper aims to develop a theoretical framework for assessing the resilience of linear systems to different types of cyber-attacks. 
A large part of the current literature on CPS security assumes complete knowledge of the system parameters, and analyzes the consequences of attacks on these systems. 
%Most present day CPS have a large number of state and measured variables. 
%System parameters in such large scale systems are prone to variations, albeit, in a small range of values. 
Parameters in CPSs with a large number of variables are prone to variations. %, albeit, in a small range of values. 
Analysis based on these models for every possible numerical realization of the system variables will be computationally infeasible. 
The structured systems approach \cite{lin1974structural} offers a way out of this conundrum. 
This technique presumes knowledge of just the zero structures (that is, the \emph{positions} of zero and nonzero entries) of the system matrices to infer system properties. 
This approach is attractive since these properties will hold for almost every valid numerical realization. 

\textbf{\emph{Contributions}}: Using linear structured system models for CPSs, we present conditions under which attacks may compromise the controllability of the system. 
%The CPS is modeled as a linear structured system, and structural conditions for an attack to be successful, in terms of disrupting or obtaining controllability of a (modified) linear structured system are provided. 
%That is, given a structured system that is structurally controllable before an attack, and a partition of the columns of the input matrix corresponding to defender and attacker nodes, the goal is to devise a strategy to ensure structural controllability in the face of an attack. 
The structural resilience to denial of service (DoS) attacks and integrity attacks is characterized in terms of the structural controllability of an associated linear structured system. 
\begin{itemize}
\item During a \emph{denial of service attack}, access to a subset of the inputs is blocked by the attacker. Our goal will be to formulate conditions for structural resilience in the absence of these inputs. 
\item An \emph{integrity attack} occurs when a state feedback strategy is not implemented appropriately. That is, only some components of the input are faithfully reproduced, while the remaining are arbitrary. 
\end{itemize}

In this light, the contributions of this paper are: 
\begin{enumerate}
\item First, we characterize the structural resilience of the system in terms of unmatched vertices in maximum matchings of the bipartite graph and connected components of the directed graph representations of the system under attack. 
\item Next, we present conditions under which a system that is already structurally resilient to a DoS attack will also be structurally resilient to a type of integrity attack called a state feedback integrity attack. 
\item Finally, we provide extensions to the case of switched linear systems (SLSs). 
SLSs are systems that can operate in one of several \emph{modes}, each of which is a linear system, and can switch from one mode of operation to another. 
We derive graph theoretic conditions for the structural resilience of such systems to DoS attacks. 
\end{enumerate}
%The attack scenarios are described in Section (\ref{Prelim}). 
%Formally, given a pre-attack configuration, $\dot{x}(t)=[A]x(t)+[B]u(t)$, with $([A],[B])$ structurally controllable, and a partition of $[B]$ corresponding to defender and attacker nodes, determine matrices $[B_2]$ and $[K_1]$ such that the post-attack system, $([A]+[B_1][K_1],[B_2])$ is structurally controllable. 
%We assume that the structure of $[B_1]$ is known.% and that all states are available for measurement. 
%The more general case when only some states or a linear combination of the states will be available for measurement is a subject for future work. 
%An additional consequence is that we are not particularly concerned about the modified system matrix containing structurally fixed modes. 
%The absence of structurally fixed modes is usually a constraint imposed while designing a control configuration to allow for arbitrary pole placement. 
%This means that no assumptions on the structural controllability of $([A],[B_1])$ is required while designing $[K_1]$. 
\subsection{Related Work}

%This section summarizes prior work in the literature that is relevant to the topics discussed in this paper. 
%
There is a large body of work that addresses modeling and detection of attacks on linear time invariant (LTI) systems. 
System and graph theoretic conditions were presented in \cite{pasqualetti2015control, pasqualetti2013attack} for an attack on a CPS (modeled as a linear descriptor system subject to unkown inputs) to be undetectable and unidentifiable by monitors.
In \cite{pajic2013topological}, for a wireless control network modeled as a discrete time linear time invariant system, under the assumption that $(A,B)$ was stabilizable and $(A,C)$ was detectable, the authors presented methods to determine a subset of columns $B_I \subset B$, and a subset of rows, $C_J \subset C$ such that $(A,B_I)$ was stabilizable and $(A,C_J)$ was detectable. %\footnote{Here, $A, B$, and $C$ are the system, input, and output matrices of a linear time invariant system: $\dot{x}=Ax+Bu$; $y=Cx$.}. 
The authors of \cite{weerakkody2017graph} and \cite{weerakkody2017robust} studied the design of distributed control systems in order to detect integrity attacks. 
They characterized the `unattackability' of a system in terms of the left-invertibility of a system matrix and strong observability of the system. 
This was extended to the structural setting by considering vertex separators, that allowed characterization of `un-attackability' from all sets of feasible malicious nodes. 

%Additionally, we assume that the structure of $B$ is known. 
The success of different kinds of attacks on LTI systems in terms of the ability to ensure or disrupt controllability of a suitably modified LTI system was characterized in \cite{barreto2013controllability}. 
We wish to extend this approach to structured linear systems. 
Interpreting security properties within this framework will allow for a characterization of resilience to attacks for general classes of CPSs. 
%rigorous analysis of more complex, networked cyberphysical systems that are vulnerable to attacks. 
Classes of attacks were also modeled using notions from game theory in \cite{barreto2013controllability}, but we do not provide an analogue in this work.
%We note that \cite{barreto2013controllability} also modeled classes of attacks using notions from game theory, but we do not provide an analogue in this work.%; instead, we quantify robustness to attacks and the worst case cost of resilience as solutions to optimization problems. 
%
%\cite{amin2009safe} provides a semidefinite programming base dsolution to solve the problem of optimally desigining a controller to minimize the impact of a class of denial of service attacks. 
%The controller thus designed satisfies certain safety specifications with high probability and power constraints in expectation. 

A survey of research on structural systems theory was recently presented in \cite{ramos2020structural}. We summarize contributions on this topic relevant to our problem in the rest of this section. 
 
The structural design of large scale systems was studied in \cite{pequito2015tac}. 
The input and output matrices were designed to select the smallest number of variables to ensure structural controllability and observability. 
The state feedback matrix was then designed to ensure the minimum number of input-output interconnections and such that the closed loop system had no structural fixed modes (so that closed loop poles can be placed arbitrarily). %\footnote{The absence of structural fixed modes will allow arbitrary placement of the closed loop poles of the system.}. 
%The design is carried out in two steps: 
%first, $[B]$ and $[C]$ are designed to $\mathrm{min}. (||[B]||_0+||[C]||_0)$ such that $([A],[B])$ is structurally controllable and $([A],[C])$ is structurally observable. 
%This is then used to design $[K]$ to $\mathrm{min}. ||K||_0$ to ensure arbitrary pole placement. 
%Here, $||\cdot||_0$ is the cardinality of the nonzero entries in the matrix.

%Control selection problems have attracted a lot of attention of late. 
For an LTI system, given $A$, the \emph{minimal controllability problem} aims to find the sparsest $B$, that will ensure that $(A,B)$ is controllable. 
In the unconstrained case, this problem was shown to be $NP-$hard in \cite{olshevsky2014minimal}. 
Interestingly, the authors of \cite{pequito2015tac} showed that the \emph{minimal structural controllability problem} was polynomially solvable. 
The minimal controllability problem for single input structural systems was studied in \cite{commault2015single}, which showed that this problem was solvable when a rank condition was satisfied. %; in the case when no structure was imposed on the system matrix, the problem was solvable with a single nonzero entry in the input matrix. 
The authors of \cite{pequito2014complexity} showed that the \emph{minimum constrained input selection problem} was $NP-$hard. 
%In this problem, given the structures of the system and input matrices, the goal was to determine a minimal set of indices of columns of the input matrix to ensure structural controllability. 
%They also showed that if the system matrix had a certain structure, the minimum dedicated input selection problem was polynomially solvable. 
%
In \cite{pequito2015minimum1}, given the costs of actuating each state, the \emph{minimum cost structural controllability problem} was shown to be polynomially solvable. 
This work was extended to the constrained case in \cite{pequito2015minimum2}, and the \emph{minimum cost constrained structural controllability problem} was shown to be $NP-$hard. % by deriving a reduction from the constrained minimum input selection problem. 
This problem was polynomially solvable when the system matrix was irreducible. % or, equivalently, the directed graph of the system was strongly connected. 
%this problem was seen to be polynomially solvable. 
%We note that most of the recent work only deals with determining the smallest subset of the $[B]$ matrix to ensure controllability of the system. 
%However, the structural controllability of the system can also be influenced by changing the number of connections from controls to states. 
%The results presented in this paper involve relating the structural resilience to an attack to the structural controllability of the system when some of its inputs are disconnected. 

Robust security indices for actuators were proposed in \cite{milovsevic2018security, milovsevic2020actuator}. The security index was defined as the minimum number of system components that had to be compromised in order to carry out a perfectly undetectable attack, and computationally efficient methods to compute the robust security index were developed. 
A security index in the form of smallest number of critical nodes to mitigate failures and ensure structural controllability was proposed in \cite{zhang2019driver}. 
The authors of \cite{alcaraz2017cyber} proposed a checkpoint-based method to verify the health of a networked control system and characterize the trustworthiness of system components. 

A parallel body of work studied in \cite{pequito2018analysis} focused on the resilience of single-mode structured systems in the face of sensor-actuator communication failures for given structured matrices [A], [B], [C]. An efficient algorithm to solve the minimum actuation-sensing-communication co-design problem under disruptive scenarios was also proposed. In comparison, our work studies structural resilience under different classes of attacks, and we additionally investigate the structural resilience of SLSs.

Structural controllability of SLSs was studied in \cite{liu2013structural}, where union graphs and colored union graphs were used to determine conditions that would ensure structural controllability. 
%In particular, a switched system can be controllable even when each of its individual modes is not controllable. 
The problem of determining the smallest subset of actuators needed to ensure structural controllability of the SLS was studied in \cite{pequito2017structural}. 
The authors also presented a polynomial algorithm to determine such a subset of actuators. 
However, the problem of selecting a minimum collection of modes from among a sequence of modes to ensure that the SLS is structurally controllable was shown to be $NP-$hard. 
%
%In this paper, we formulate conditions to ensure the structural resilience of a system to an attack in relation to the structural controllability of the system after a subset of its inputs are `disconnected'. 
%\textbf{show example where system is $(U_{s_1} \cup U_{s_2} \cup U_{att})-$ controllable, $(U_{s_1} \cup U_{att})-$ controllable, but not $(U_{s_1} \cup U_{s_2})-$ controllable}.
%This means that an adversary can affect controllability of the system by \emph{flooding} the (particular nodes of the) network with control signals. 
\subsection{Outline of Paper}

Section \ref{Prelim} is a primer on linear structured systems and graph theory. 
Section \ref{Prob} states the problem to be solved, and summarizes some existing results on structural controllability. 
The main results of this paper are presented in Sections \ref{DoSRes} and \ref{IntegRes}. %, \ref{SFIntegRes}). 
%Section (\ref{Cost}) formulates optimization problems for the cases when the defender is robust to an arbitrary attack, and when an attacker is not able to disrupt controllability, but wants to ensure that the defender incurs maximum cost in maintaining resilience to an attack, while keeping its own cost low. 
%The computational complexity of the results only rely on determining maximum matchings in bipartite graphs and strongly connected components of directed graphs. 
Section \ref{complex} makes a note of the computational complexity of the results. 
Section \ref{Examples} presents illustrative examples. 
We characterize the structural resilience of SLSs to DoS attacks in Section \ref{Switched}. 
We conclude by presenting possible directions for future research in Section \ref{Future}. 

This paper is different from a preliminary version that appeared in \cite{bhaskar2016structres} in the following ways: 
(i) we provide complete proofs of all results- most notably, for Propositions \ref{DoSThm} and \ref{DoSThm1}, and Theorem \ref{ThmInteg}, 
(ii) we introduce a notion of \emph{complete controllability}, and prove a related result in Theorem \ref{CorInteg}, 
(iii) we provide a characterization of the structural resilience of switched linear systems to DoS attacks, and 
(iv) we present a discussion on the computational complexity of our approach. 
We additionally incorporate clarifying text throughout the paper to improve readability. 
\section{Preliminaries} \label{Prelim}

This section gives an introduction to structured linear systems and graph theory. %, and defines several terms that will be needed to understand the results in this paper. 
A more detailed exposition and references to prior work in the area can be found in \cite{dion2003generic}.% for a more detalied exposition and references to prior work in this area. 
\subsection{Structured Linear Systems}

Consider an LTI system: 
$\dot{x}(t)=Ax(t)+Bu(t)$, with $x(t) \in \bb{R}^n$, $u(t) \in \bb{R}^p$, $A \in \bb{R}^{n \times n}$ and $B \in \bb{R}^{n \times p}$. 

\begin{df}
The LTI system is \emph{controllable} if for every initial state $x(0)$ and final state $x(t_f)$, there exists an input $u(\cdot)$ on $[0,t_f]$ that transfers the system from $x(0)$ to $x(t_f)$. 
\end{df}

%Verifying the controllability of an LTI system of the form in equation (\ref{LinSysCont}) where the states and inputs are defined on finite dimensional vector spaces is equivalent to checking a matrix rank condition, as stated in the following result. 
%
\begin{thm} \cite{rugh1996linear}
The LTI system is controllable if and only if $rank(\left[\begin{matrix}
B&AB&\dots&A^{n-1}B
\end{matrix} \right])=n$. 
\end{thm}

The structural systems framework \cite{lin1974structural} assumes knowledge of only the zero stuctures, $[A] \in \{0,*\}^{n \times n}$ and $[B] \in \{0,*\}^{n \times p}$, of $A$ and $B$ respectively. 
That is, every entry in $[A]$ and $[B]$ is either a \emph{fixed} zero or a \emph{free} parameter (which can take any numerical value, including $0$). 
$[A]$ and $[B]$ are called \emph{structured matrices}. 
The rows and columns of $[A]$ indicate how the states of the system influence one another. 
A nonzero entry $a_{ij} \in [A]$ indicates that the $j^{th}$ component of the state vector, $x_j$, influences changes in the $i^{th}$ component, $x_i$ (the $j^{th}$ and $i^{th}$ entries in the state vector of dimension $n$).  
The rows and columns of $[B]$ indicate how inputs to the system influence the states. 
A nonzero entry $b_{ij} \in [B]$ indicates that a change in $x_i$ is influenced by the input $u_j$ (the $j^{th}$ entry in the input vector of dimension $p$). 
A zero entry would imply the lack of an interconnection between corresponding variables. 
One can think of the structured representation of a system in the following way:
\begin{eg} 
Consider a symmetric structured matrix $[H] \in \{0,*\}^{n \times n}$ representing a power system. 
The dimension of $[H]$, $n$, is indicative of the number of components in the system (generators, transformers, loads). 
$h_{ij}=*$ signifies that there is a wire connecting components $i$ and $j$, with the direction of current through the wire from $j$ to $i$. 
A fixed zero entry in $[H]$ corresponds to the absence of a wire between the respective components. 
$h_{ij}=*$ is an indication that changes in the numerical value of a parameter associated with component $j$ influences changes in the numerical value of a parameter associated with component $i$. 
This parameter could be the current flowing through the component, or the voltage drop across the component, and is not precluded from being set to (the numerical value) zero. 
For example, when two purely resistive loads are connected to each other, $h_{ij}=h_{ji}=*$ %; in this case, the free parameter $*$ 
will take the numerical value $0$ when both loads are isolated from a source. 
\end{eg}

%\begin{rem}
%%In the sequel, we will use the terms \emph{state vertices} and \emph{input vertices} to refer to the appropriate rows or columns of $[A]$ and $[B]$ in the structural setting. 
%In the sequel, the components of the \emph{state (input) vectors} will correspond to \emph{state (input) vertices} in a directed graph. 
%As we will describe in the next part of this section, the edges in this graph will be determined by the $[A]$ and $[B]$ matrices. 
%\end{rem}
%
A matrix $H\in \bb{R}^{m \times n}$ with the same zero structure as the structured matrix $[H]\in \{0,*\}^{m \times n}$ is called an admissible numerical realization (ANR) of $[H]$. 
The structural representation of a system will enable the analysis of system properties in a \emph{generic} sense. That is, the set of values of parameters for which a property will not hold will be a set of Lebesgue measure zero \cite{dion2003generic}. As a consequence, the property will hold for \emph{almost} every ANR. 
\begin{df}
$([A],[B])$ is \emph{structurally controllable} if there exists an ANR $(A,B)$ that is controllable. 
\end{df}
\begin{rem}
If $([A],[B])$ is structurally controllable, then almost every ANR will be controllable\footnote{Some authors refer to such a system as \emph{generically controllable} \cite{dion2003generic}}. 
\end{rem}

\subsection{Graph Theory}

Directed graphs (digraphs) provide an elegant means to represent linear structured systems \cite{pequito2015tac}. 
%Many system theoretic properties can be analyzed and deduced from an equivalent representation as a (directed) graph. 
Properties of the system such as controllability and observability can be inferred from the digraph associated with the system, and independently of numerical values of parameters. 
This makes it an attractive tool to study large scale, complex systems, on which performing computations using numerical values of variables will invariably be costly. 
Consider the linear structured system 
$\dot {x}(t)=[A]x(t)+[B]u(t)$, where, $x(t) \in \bb{R}^n$, $u(t) \in \bb{R}^p$, %$y \in \ca{Y} \subseteq \bb{R}^m$, 
$[A] \in \{0, *\}^{n \times n}$ and $[B] \in \{0,*\}^{n \times p}$. %, $[C] \in \{0,*\}^{m \times n}$ and $[D] \in \{0,*\}^{m \times p}$.

The \emph{directed graph} of the structured system is $\ca{D}=(\ca{V}, \ca{E})$, where $\ca{V}=\{u_1,\dots,u_m,x_1,\dots,x_n\}:=\{\ca{U},\ca{X}\}$ and $\ca{E}=\ca{E}_A \cup \ca{E}_B$, 
% \cup \ca{E}_C \cup \ca{E}_D$
where $\ca{E}_A=\{(x_j,x_i)| [A]_{ij} \neq 0\}$, $\ca{E}_B=\{(u_j,x_i)| [B]_{ij} \neq 0\}$.%, $\ca{E}_C=\{(x_j,y_i)| [C]_{ij} \neq 0\}$, $\ca{E}_D=\{(u_j,y_i)| [D]_{ij} \neq 0\}$. 

A sequence of directed edges $\{(v_1,v_2),(v_2,v_3),\dots,(v_{k-1},v_k)\}$ is a \emph{simple path} from $v_1$ to $v_k$ if $v_1,\dots,v_k$ are all distinct. 
The simple path $\{(v_1,v_2),(v_2,v_3),\dots,(v_{k-1},v_k)\}$, with an additional edge, $(v_k,v_1)$, or a vertex with a self loop, is called a \emph{cycle}. 
A vertex $v'_2$ is \emph{reachable} from another vertex $v'_1$ if there exists a simple path from $v'_1$ to $v'_2$. 
Let $\ca{V}_1, \ca{V}_2 \subseteq \ca{V}$. 
Two paths from $\ca{V}_1$ to $\ca{V}_2$ are \emph{disjoint} if they consist of  disjoint sets of vertices. 
A set of $v$ mutually disjoint and simple paths from $\ca{V}_1$ to $\ca{V}_2$ is a \emph{linking} of size $v$ from $\ca{V}_1$ to $\ca{V}_2$. 
A \emph{cycle family} is a set of mutually disjoint cycles. 
A \emph{$\ca{U}-$rooted path} is a simple path with source vertex in $\ca{U}$. 
A \emph{$\ca{U}-$rooted path family} is a set of mutually disjoint $\ca{U}-$rooted paths. 

A digraph $\ca{D}_s=(\ca{V}_s,\ca{E}_s)$ is a \emph{subgraph} of $\ca{D}$ if $\ca{V}_s \subseteq \ca{V}$ and $\ca{E}_s \subseteq \ca{E}$. 
%If $\ca{V}_s=\ca{V}$, then $\ca{D}_s$ is said to \emph{span} $\ca{D}$. 
A subgraph $\ca{D}_s$ satisfying a property $P$ is \emph{maximal} if there is no other subgraph $\ca{D}_{s'}$ such that $\ca{D}_s$ is a strict subgraph\footnote{A subgraph is \emph{strict} if at least one of  $\ca{V}_s \subset \ca{V}$ or $\ca{E}_s \subset \ca{E}$ holds.} of $\ca{D}_{s'}$ and property $P$ holds for $\ca{D}_{s'}$. 

$\ca{D}$ is \emph{strongly connected} if there is a simple path from each vertex to every other vertex in the graph. 
A \emph{strongly connected component (SCC)} is a maximal subgraph $\ca{D}_S$, of $\ca{D}$, such that $\ca{D}_S$ is strongly connected. 
With SCCs as supernodes, one can generate a \emph{directed acyclic graph (DAG)} in which each supernode corresponds to an SCC, and there exists a directed edge from one SCC to another if and only if there exists an edge from a node in the first SCC to some node in the second SCC in the original graph. 
%there exists a directed edge between two SCCs if and only if there exists a directed edge connecting vertices in the SCCs in the original digraph. 
An SCC is \emph{linked} if it has at least one incoming (outgoing) edge to (from) its vertices from (to) vertices of another SCC. 
An SCC is \emph{non top linked} if it has no incoming edges to its vertices from vertices of another SCC\footnote{Non top linked SCCs are called \emph{source} SCCs in the graph theory literature. In this paper, we will use the terminology from \cite{pequito2015tac}.}.%\footnote{This notion relies on the assumption that the directed graph is drawn in a way that the edges between SCCs point down, as in Figure (\ref{fig0a})}.% and \emph{non bottom linked} if it has no outgoing edges. 

A \emph{bipartite graph}, denoted $\ca{B}(\ca{V}_1,\ca{V}_2,\ca{E}_{\ca{V}_1,\ca{V}_2})$, is a graph whose vertices can be divided into disjoint sets $\ca{V}_1$ and $\ca{V}_2$ such that every edge in the graph is from a vertex in $\ca{V}_1$ to a vertex in $\ca{V}_2$, or from a vertex in $\ca{V}_2$ to a vertex in $\ca{V}_1$. 
%The bipartite graph is denoted $\ca{B}(\ca{V}_1,\ca{V}_2,\ca{E}_{\ca{V}_1,\ca{V}_2})$. 
In this paper, we will restrict our discussion to bipartite graphs in which all edges are directed from $\ca{V}_1$ to $\ca{V}_2$, that is, $\ca{E}_{\ca{V}_1,\ca{V}_2} \subset \{(v_1,v_2)|v_1 \in \ca{V}_1, v_2 \in \ca{V}_2\}$. 
%
%For any $\ca{V}_1, \ca{V}_2$, a \emph{bipartite graph} $\ca{B}(\ca{V}_1,\ca{V}_2,\ca{E}_{\ca{V}_1,\ca{V}_2})$ is a digraph with vertex set $\ca{V}_1 \cup \ca{V}_2$ and edge set $\ca{E}_{\ca{V}_1,\ca{V}_2} \subset \{(v_1,v_2)|v_1 \in \ca{V}_1, v_2 \in \ca{V}_2\}$\footnote{This is a slightly stronger condition than the general bipartite case, where one can have directed edges $\ca{V}_2$ to $\ca{V}_1$ as well.}. 
$\ca{B}(\ca{V}_1,\ca{V}_2,\ca{E}_{\ca{V}_1,\ca{V}_2})$ can also be associated with a matrix $H$ with $|\ca{V}_1|$ columns and $|\ca{V}_2|$ rows, with $\ca{E}_{\ca{V}_1,\ca{V}_2} = \{(v_{1_j},v_{2_i}): [H]_{ij} \neq 0\}$. 
Given $\ca{B}(\ca{V}_1,\ca{V}_2,\ca{E}_{\ca{V}_1,\ca{V}_2})$, a \emph{matching} is a subset of edges that do not share vertices. 
A \emph{maximum matching} is a matching that has the largest number of edges. 
Vertices not belonging to a maximum matching are called \emph{unmatched}. 
An unmatched vertex $v_2 \in \ca{V}_2$ (respectively, $v_1 \in \ca{V}_1$) is called a \emph{right unmatched vertex} (\emph{left unmatched vertex}). 
%If a maximum matching has no unmatched vertices, then it is called a \emph{perfect matching}. 
A \emph{perfect matching} is  a maximum matching with no unmatched vertices. 

The \emph{bipartite graph associated with a directed graph} $\ca{D}(\ca{V},\ca{E})$ is constructed as follows \cite{brualdi1980bigraphs}: to each $v_i \in \ca{V}$, we associate two vertices $s_i$ and $w_i$. 
There is a directed edge from $s_i$ to $w_j$ in the new graph if and only if there is an edge from $v_i$ to $v_j$ in $\ca{D}(\ca{V},\ca{E})$. 
We abuse notation by using $\ca{B}(\ca{V},\ca{V},\ca{E})$ to denote the bipartite graph associated with $\ca{D}(\ca{V},\ca{E})$. 

A \emph{top assignable SCC} of $\ca{D}([A])=(\ca{X},\ca{E}_A)$ is a non-top-linked SCC which contains at least one right unmatched vertex in a maximum matching. Since a maximum matching is not unique, whether an SCC is top assignable will depend on the maximum matching under consideration. 
The \emph{maximum top assignability index} of $\ca{D}([A])$ is the maximum number of top assignable SCCs among the maximum matchings associated with $\ca{B}([A])$. 
\begin{eg}\label{matrix}
Figure (\ref{fig0a}) shows the directed and bipartite graph representations of a matrix $[A]$ given below: 
\begin{align}
[A]&=
\begin{bmatrix}
0 & 0 & * & 0 & 0 & 0 & 0  \\
* & 0 & 0 & 0 & 0 & 0 & 0 \\
0 & * & 0 & 0 & 0 & 0 & 0 \\
0 & * & 0 & 0 & 0 & 0 & 0 \\
0 & 0 & 0 & 0 & 0 & 0 & 0  \\
0 & 0 & 0 & * & 0 & 0 & *  \\
0 & 0 & 0 & 0 & * & * & 0 \nonumber
\end{bmatrix}
\end{align}
\begin{figure}
 \centering
  \includegraphics[width=1.7 in]{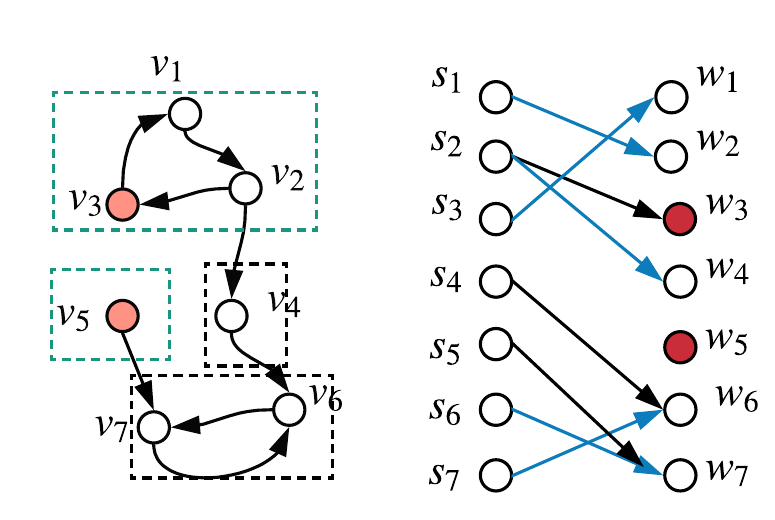} 
\caption{Structured system of Example  \ref{matrix} as a graph} \label{fig0a} 
\end{figure}

The SCCs of the directed graph, $\ca{D}([A])$, are the vertices within each dotted box. 
The dotted boxes in green (comprising the vertex $(v_5)$ and the vertices $(v_1,v_2,v_3)$) represent the non top-linked SCCs. 
The bipartite graph representation, $\ca{B}([A])$ is got by duplicating each vertex of the directed graph, and the edges are determined by the edges in $\ca{D}([A])$. 
The edges of $\ca{B}([A])$ in blue form a maximum matching. 
Removing the vertices that are incident on edges in the maximum matching, we see that $w_3$ and $w_5$ are right unmatched vertices. 
We see that $w_3$ and $w_5$ in $\ca{B}([A])$ correspond to $v_3$ and $v_5$ in $\ca{D}([A])$, which both belong to non-top linked SCCs, which makes these SCCs top-assignable. 

Notice that this maximum matching is not unique. 
Another maximum matching could be got by removing the edge $(s_2 \rightarrow w_4)$ from the previous maximum matching and adding the edge $(s_2 \rightarrow w_3)$. 
The right unmatched vertices of this maximum matching will be $w_4$ and $w_5$. 
\end{eg}
\section{Problem Formulation} \label{Prob}

Removing the explicit dependence on $t$, and rewriting $u(t)$ as $u=\bigl(\begin{matrix}
u_{1}&\dots&u_d&u_{d+1}&\dots&u_{p}
\end{matrix} \bigr)^T$, 
we will use $u_{def} \in \bb{R}^d$ and $u_{att}\in \bb{R}^a$ (with $a:=p-d$) to collectively denote the elements $\bigl(\begin{matrix} u_{1}&\dots&u_d\end{matrix} \bigr)^T$ and $\bigl(\begin{matrix} u_{d+1}&\dots&u_{p} \end{matrix} \bigr)^T$ respectively. 
The sets $u_{def}$ and $u_{att}$ represent the input vertices accessible to the system (defender) and attacker respectively. 
The structural resilience of the system to the different types of attacks discussed in this paper will depend, to a large extent, on the cardinality of the vertex sets $u_{def}$ and $u_{att}$ (that is, on $d$ and $a$) \emph{vis-\`a-vis} the number of unmatched state vertices. 
The system model is now:
\begin{align}
\dot{x}(t)&=[A]x(t)+[B_{def}]u_{def}(t)+[B_{att}]u_{att}(t) \label{SysModel}
\end{align}

Define $\ca{X}_{def}:=\{x| u_i \rightarrow x \text{ for some } i \in \{1,\dots,d\}\}$ and $\ca{X}_{att}:=\{x|u_j \rightarrow x \text{ for some } j \in \{d+1,\dots,p\}\}$. 
These are the sets of state vertices that can be directly connected to inputs controlled by the defender and attacker respectively. 
\begin{ass}\label{UXConn}
$\ca{X}_{def}$ and $\ca{X}_{att}$ are disjoint. 
\end{ass}

This is a reasonable assumption in that it means that the defender (system) will have (limited) access to only a subset of the state vertices which it can `directly' control ($\ca{X}_{def}$) in order to be resilient to an attack. 
Once the attacker has gained access to the system by manipulating a subset of the inputs, thereby influencing a set of states ($\ca{X}_{att}$), it retains access to these states while the defender tries to ensure that the system is resilient to the attack by appropriately controlling the other states ($\ca{X}_{def}$). 

Assumption \ref{UXConn} can also be viewed in light of the setting where inputs in $u_{def}$ (and consequently, states in $\ca{X}_{def}$) are deemed to be `trustworthy', in the sense that they cannot be tampered with. 
Our results then seek to determine conditions on $|\ca{X}_{def}|$ in order to ensure structural resilience. 
We note that the defender does not need to have knowledge of which states the adversary might be able to influence- in the worst case, $\ca{X}_{att} = \ca{X} \setminus \ca{X}_{def}$, where $\ca{X}$ denotes the complete set of state vertices. However, our results will only require Assumption \ref{UXConn}, which is less restrictive. 

In the structural setting, this would imply that $[B_{def}]$ will have fixed zeros in rows corresponding to $\ca{X}_{att}$, and $[B_{att}]$ will have fixed zeros in rows corresponding to $\ca{X}_{def}$. 
Specifically, the only possibly non-zero entries in $[B_{def}]$ will be in rows that correspond to states in $\ca{X}_{def}$, and the only possibly non-zero entries in $[B_{att}]$ will be in rows that correspond to states in $\ca{X}_{att}$. 
 
The resilience of the CPS will be characterized in terms of the structural controllability of the system when it is subject to an attack. 
This will subsequently be shown to be equivalent to formulating conditions on the non-attacked nodes in the graph of the structured system. 
Throughout this paper, we shall assume that the sets $\ca{X}_{def}$ and $\ca{X}_{att}$ remain unchanged with time. 
The system will be structurally resilient to an attack if it is structurally controllable when it has `access' to only some components of the state vector, while the remaining components of the state vector (those under `attack') cannot be directly accessed by it. 
While this is a conservative assumption, considering the scenario when the set of compromised nodes varies with time is an interesting problem that we will consider in future work. 

At this juncture, we would like to point out two different ways of viewing a DoS attack. 
In the cybersecurity literature, a DoS attack typically occurs when an adversary `floods' the system with spurious inputs or requests, thereby ensuring that the system cannot address `genuine' service requests. 
In our framework, however, we view a DoS attack in terms of ensuring the structural resilience of the system when certain inputs (corresponding to the attacker) are disregarded. 
A spurious input is assumed to not be of use, and is therefore set to zero. 
We then want to see if the system can satisfy certain properties in order to be structurally resilient in the absence of these inputs\footnote{The distinction between an attack and a fault is somewhat arbitrary, especially if only one input is compromised. When $>1$ input is compromised, this is more likely evidence of an attack than a fault. However, from the standpoint of the analysis in this paper, in both cases, our goal is to characterize when the system remains controllable, even with some compromised inputs. Of course, an engineer charged with redesigning a system that is not resilient to compromised inputs will need to know if the problem is due to faulty components (eg. bad sensors) or an attacker. Our point is that such an engineer should consider both possibilities.}. 
We formally state the problem that we wish to solve:
\begin{prob} \label{StructRes}
Given the system (\ref{SysModel}) with $([A],[B])$ structurally controllable before an attack, characterize its structural resilience to denial of service (DoS) and integrity attacks.
\end{prob}

The next three results present conditions for structural controllability, and lower bounds on the number of control inputs and input to state links. 
We leverage the insight from these results to characterize the resilience when an attacker may influence certain inputs and/ or states of the structured system.  
The reader is directed to the references cited for complete proofs of these results. 
\begin{thm} \cite{pequito2015tac,dion2003generic}\label{StructContThm}
The following are equivalent:
\begin{enumerate}
\item $([A],[B])$ is structurally controllable. 
%\item $\ca{D}([A],[B]):=(\ca{X} \cup \ca{U}, \ca{E}_A \cup \ca{E}_B)$ is spanned by a disjoint union of input cacti. 
\item Every state vertex is the end of a $\ca{U}-$rooted path and there exists a union of a $\ca{U}-$rooted path family and a cycle family containing all vertices in $\ca{X}$. 
\item Every right unmatched vertex of a maximum matching of $\ca{B}([A],[B])$ is connected to a distinct input, and one state vertex from each non-top-linked SCC of $\ca{D}([A])$ is connected to some input. 
\end{enumerate}
\end{thm}
\begin{thm} \cite{liu2011controllability}\label{MinInp}
Let $m$ be the number of right unmatched vertices in a maximum matching of $\ca{B}([A])$. 
Then, the minimum number of inputs needed to ensure structural controllability is one, if $m=0$, and $m$, otherwise.
\end{thm}
\begin{thm}\cite{pequito2015tac}\label{MinLinks}
Let $\beta$ be the number of non-top-linked SCCs and $\alpha$ the maximum top assignability index in $\ca{D}([A])$. 
Then, the minimum number of input-state links needed to ensure structural controllability is $m+\beta-\alpha$. 
\end{thm}

From the above results, we observe that one way to reduce the number of input to state links needed to ensure structural controllability is to determine a maximum matching of $\ca{B}([A])$ such that as many right unmatched vertices belong to non-top linked SCCs. 
This will ensure that $\beta - \alpha$ is `close' to zero, and the minimum number of input to state links needed is `close' to $m$, the number of right unmatched vertices. In the sequel, we assume $m > 0$. 

We conclude this section by defining what it means for an attack to be structurally successful. 
The system post-attack is defined to be the configuration for which structural controllability has to be ensured when only vertices in $\ca{X}_{def}$ can be connected to inputs.
\begin{df}\label{StructSucc}
An attack on the system is said to be \emph{structurally successful} if the system post-attack is not structurally controllable. 
The system is \emph{structurally resilient} to the attack if the system post-attack is structurally controllable. 
\end{df} 
\section{Structural Resilience to DoS Attacks} \label{DoSRes}
This section presents our main results. 
We characterize the resilience of a structured system to denial-of-service (DoS) attacks in terms of certain properties inherent to a graph-theoretic representation of the system. 

During a DoS attack, the attacker blocks access to inputs in $u_{att}$. 
The system still has access to inputs in $u_{def}$. 
%Thus, while $u_{att}$ is zero, $u_{def}$ is arbitrary. 
Structurally, this corresponds to determining the matrix $[B_{def}]$, with $[B_{att}]=0$, to ensure structural resilience. 
The system model is given by:
\begin{align} 
\dot{x}(t)&=[A]x(t)+[B_{def}]u_{def}(t) \label{DoSModel}
\end{align}
 
Let $m_{def}$ ($m_{att}$) be the number of right unmatched vertices in $\ca{B}([A])$ corresponding to $\ca{X}_{def}$ ($\ca{X}_{att}$). %Further, let $m > 0$.
$l(P \rightarrow Q)$ denotes the set of links from $P$ to $Q$. 
Proposition \ref{DoSThm} provides a sufficient condition for a DoS attack to be successful. 
%The proof is omitted for brevity. 
\begin{pr} \label{DoSThm}
A DoS attack on the system in (\ref{SysModel}) is structurally successful if:
\begin{enumerate} \label{USUA}
\item $p \geq m+\beta-\alpha$, (where $m_{def}+m_{att}=m$) \emph{OR} 
\item $p \geq m$ and $|l(u \rightarrow \ca{X})| \geq m+\beta-\alpha$
\end{enumerate}
and $d < m_{def}$, where $p$ ($d$) is the dimension of $u$ ($u_{def}$).
\end{pr}
\begin{IEEEproof}$([A],[B])$ is assumed to be structurally controllable before an attack occurs. 
This means that there are at least $m$ vertices in $u$ and $m+\beta-\alpha$ links from $u$ to $\ca{X}$, which gives the inequalities in $1)$ and $2)$. % in (\ref{USUA}). 
The last inequality is obtained from the fact that if, after an attack, the number of available inputs is less than the number of right unmatched vertices in $\ca{B}([A])$ corresponding to $\ca{X}_{def}$, then $([A],[B_{def}])$ will not be structurally controllable.
Thus, the system will not be able to mitigate the effect of the attack.
\end{IEEEproof}

The conditions of Proposition \ref{DoSThm} are not necessary- an attack could be successful even when $p \geq m+\beta-\alpha$ and $d \geq m_{def}$. 
Although the minimum input requirement is satisfied, the conditions to ensure structural controllability must be carefully checked. 
\begin{pr} \label{DoSThm1}
If $d \geq m_{def}$, a DoS attack is structurally successful if:
\begin{enumerate}
\item There is an unreachable state from vertices of $u_{def}$. OR
\item There does not exist a disjoint union of $u_{def}$ rooted path families and cycle families covering all the states. OR 
\item $|l(u_{def} \rightarrow \ca{X})| < m_{def}+\beta-\alpha$. OR 
\item Every maximum matching of $\ca{B}([A])$ has a right unmatched vertex in $\ca{X}_{att}$. OR 
\item There is a non-top-linked SCC in $\ca{D}([A])$ comprising only vertices from $\ca{X}_{att}$. 
\end{enumerate}
\end{pr}
\begin{IEEEproof}The first three conditions follow from Theorem \ref{StructContThm} and Theorem \ref{MinLinks}. 
The last two follow from the fact that inputs from $u_{def}$ cannot be assigned to vertices in $\ca{X}_{att}$. 
\end{IEEEproof}

Propositions \ref{DoSThm} and \ref{DoSThm1} together lead to the main result of this section:% the following result:
\begin{thm} \label{DoSThmMain}
Given $[A]$ and the indices of $[B]$ corresponding to $[B_{def}]$, the system in (\ref{DoSModel}) is \emph{structurally resilient to a DoS attack} if and only if $([A],[B_{def}])$ is structurally controllable and: 
\begin{enumerate}
\item there exists a maximum matching of $\ca{B}([A])$ that does not contain a right unmatched vertex in $\ca{X}_{att}$; 
\item $\mathcal{D}([A])$ does not have a non-top linked SCC comprising vertices from only $\ca{X}_{att}$.
\end{enumerate}
\end{thm}
\begin{IEEEproof}If $([A],[B_{def}])$ is not structurally controllable, then at least one of the first two conditions of Lemma \ref{DoSThm1} will not be satisfied, and the system will not be structurally resilient to a DoS attack. 

Now, let $([A],[B_{def}])$ be structurally controllable. 
Any right unmatched vertex in $\ca{X}_{att}$ or a non-top-linked SCC consisting of only vertices in $\ca{X}_{att}$ will have to be assigned to a control in $u_{def}$. 
This would violate the assumption that $u_{def}$ can only be connected to states in $\ca{X}_{def}$. 
This means that the system will not be structurally resilient to a DoS attack. 
If $([A],[B_{def}])$ is structurally controllable, the absence of right unmatched vertices or non-top-linked SCCs comprised exclusively of vertices from $\ca{X}_{att}$ corresponds to the existence of a control configuration such that $d \geq m_{def}$ and $|l(u_{def} \rightarrow \ca{X}_{def})| \geq m_{def}+\beta-\alpha$, which ensures structural resilience to a DoS attack.
\end{IEEEproof}
\begin{rem}
This is different from the minimal controllability problem, where, given $[A]$, we need to find the sparsest $[B]$ such that $([A],[B])$ is structurally controllable. 
In our framework, if the number of columns of $[B_{def}]$ exceeds a certain threshold ($m$), then the only remaining task is to fill in the `missing links' to ensure structural controllability. 
Conversely, structural controllability cannot be achieved if the number of columns of $[B_{def}]$ is below this threshold. 
\end{rem}

The results in this section establish that structural resilience to a DoS attack is intimately linked to the ability to reach every vertex in $\ca{X}_{att}$ along a directed path in $\mathcal{D}([A])$ through a control in $u_{def}$ connected to some state in $\ca{X}_{def}$. 
This ensures that states of the system can be controlled exclusively through controls in $u_{def}$ even when an attacker may block certain inputs. 

\section{Structural Resilience to Integrity Attacks} \label{IntegRes}

The previous section characterized structural resilience when an attacker disables or blocks certain inputs. 
However, in certain cases, it might be possible for the attacker to additionally influence modification of the structural representation of the system matrix $[A]$. 
One way by which this can be accomplished is through state feedback. 

State feedback is a popular control strategy in which the closed-loop poles of a system can be `placed' in order to control the characteristics of the response of the system. 
The control input is given by $u(t):=Kx(t)$, and if the system is controllable, then the eigenvalues of the modified system matrix $(A+BK)$ (called closed-loop poles) can be arbitrarily placed. 

This section characterizes the resilience of a structured system in two scenarios involving state-feedback. 
In the first, only a part of the feedback is correctly reproduced. 
In the second case, the attacker will have the ability to directly gain access to a state- this would mean the ability to add or remove certain edges to the system matrix $[A]$. 
We present each scenario in detail in the remainder of this section. 

During an integrity attack, only the part of the input corresponding to $u_{def}$ is faithfully reproduced, while that corresponding to $u_{att}$ is arbitrary. 
The attacker is deemed to be successful if the system is structurally controllable without needing to connect inputs to $\ca{X}_{def}$. 
With $[A_{def}]:=([A]+[B_{def}][K_{def}])$, the system model is:
\begin{align}
%\dot{x}(t)&=([A]+[B_{def}][K_{def}])x(t)+[B_{att}]u_{att}(t) \nonumber \\
\dot{x}(t)&=[A_{def}]x(t)+[B_{att}]u_{att}(t) \label{IntegModel}
\end{align}

\begin{rem}
We note that in contrast to Definition \ref{StructSucc}, resilience in this setting relies on the ability to connect inputs to $\ca{X}_{att}$, and not $\ca{X}_{def}$.  
\end{rem}

The following result characterizes the structural resilience of the system when it is subject to an integrity attack. 
\begin{thm}\label{ThmInteg}
The system in Equation (\ref{IntegModel}) is \emph{structurally resilient to an integrity attack} if and only if there is a right unmatched vertex in $\ca{X}_{def}$ in every maximum matching of $\ca{B}([A_{def}])$ or there exists a non-top-linked SCC of $\ca{D}([A_{def}])$ comprising exclusively vertices in $\ca{X}_{def}$.
\end{thm}
\begin{IEEEproof}
This follows from Assumption \ref{UXConn}. 
The attacker will not be able to ensure structural controllability of (\ref{IntegModel}) if some vertex in $\ca{X}_{def}$ has to be assigned to a control in $u_{att}$. 
\end{IEEEproof}

Theorem \ref{ThmInteg} studies the scenario when the system is reslient to an integrity attack as a consequence of the attacker not being able to ensure structural controllability. 
Theorem \ref{CorInteg} addresses the case when a malicious adversary could completely take over operation of the system. 
We first introduce a notion of \emph{complete controllability}. 
\begin{df}
The system in Equation (\ref{IntegModel}) is \emph{completely controllable by an attacker} if structural controllability can be achieved by only using inputs from $u_{att}$. 
\end{df}
\begin{thm}\label{CorInteg}
Completely controllablity by an attacker is possible if and only if there is at least one maximum matching of $\ca{B}([A_{def}])$ comprised exclusively of vertices from $\ca{X}_{att}$ and all non-top-linked SCCs of $\ca{D}([A_{def}])$ have vertices exclusively in $\ca{X}_{att}$.
\end{thm}
\begin{IEEEproof}
This result follows from the fact that if all vertices to which inputs have to be connected to ensure structural controllability are in $\ca{X}_{att}$, then the attacker can control the system. 
As a consequence, the system will not be structurally resilient.  
\end{IEEEproof}
%\subsection{Structural Resilience to State Feedback Integrity Attacks} \label{SFIntegRes}

Alternatively, through a measurement or other means (e.g. changing a controller parameter), an attacker might gain access to a state. 
We term this scenario a \textbf{\emph{state feedback integrity (SFI) attack}}. 
In this case, $u_{att}(t)=K_{att}x(t)$, while $u_{def}$ is arbitrary. 
For structural systems, this corresponds to designing $[B_{def}]$ to ensure structural controllability. 
With $[A_{att}]:=([A]+[B_{att}][K_{att}])$, we have:
\begin{align}
%\dot{x}(t)&=([A]+[B_{att}][K_{att}])x(t)+[B_{def}]u_{def}(t) \nonumber \\
\dot{x}(t)&=[A_{att}]x(t)+[B_{def}]u_{def}(t) \label{SFIntegModel} 
\end{align}

%The system will be structurally resilient if the attacker is unable to disrupt structural controllability.
Let $m_A$ and $m_{A_{att}}$ denote the number of right unmatched vertices in a maximum matching of $\ca{B}([A)]$ and $\ca{B}([A_{att}])$. 
Let $\ca{Z}(H)$ denote the \emph{zero structure} of a structured matrix $H$. 
A zero structure is therefore a particular configuration of $0$s and $*$s. 
For structured matrices $H$ and $H'$ of the same dimension, we write $\ca{Z}(H') \subseteq \ca{Z}(H)$  whenever $h'_{ij} = 0$ in $[H']$ implies $h_{ij}=0$ in $[H]$. 

The next result of this section provides certain guarantees on the structural resilience of the system to an SFI attack depending on its resilience to a DoS attack \cite{bhaskar2016structres}. 

\begin{thm}\label{ThmDoSToInteg}
If the system in Equation (\ref{SysModel}) is structurally resilient to a DoS attack for some $[B_{def}]$ with zero structure $\ca{Z}(B_{def})$, then there exists a $[B'_{def}]$ which satisfies $\ca{Z}(B'_{def}) \subseteq \ca{Z}(B_{def})$ for which it will also be structurally resilient to a state feedback integrity attack. 
Moreover, if 
\begin{align}
m_{A_{att}}+\beta_{A_{att}}-\alpha_{A_{att}} &\leq m_A+\beta_A - \alpha_A \label{DoSToInteg}
\end{align}
for some $[B_{def}]$ corresponding to the DoS case, then the same $[B_{def}]$ will ensure structural resilience to a state feedback integrity attack ($m, \beta$, and $\alpha$ are as in Theorem (\ref{MinLinks})).
\end{thm} 
\begin{IEEEproof}
Addition of edges corresponding to $[B_{att}][K_{att}]$ to $[A]$ will ensure that the number of right unmatched vertices in a maximum matching of $[A_{att}]$ can only be as many as the number of right unmatched vertices in a maximum matching of $[A]$.
Therefore, $m_{A_{att}} \leq m_A$. 
From Theorem \ref{MinInp} and equation (\ref{DoSModel}), structural resilience to a DoS attack implies $d \geq m_A$ holds. 
This gives $d \geq m_{A_{att}}$. 

If the inequality (\ref{DoSToInteg}) holds, then $|l(u_{def} \rightarrow \ca{X})| \geq m_{A_{att}}+\beta_{A_{att}}-\alpha_{A_{att}}$, and no additional links between inputs and states will be needed to ensure structural controllability, and $[B'_{def}]=[B_{def}]$.
Additional links will be needed if (\ref{DoSToInteg}) does not hold. 
This corresponds to adding free parameters to $[B_{def}]$, giving $[B'_{def}]$, which satisfies $\ca{Z}(B'_{def}) \subseteq \ca{Z}(B_{def})$. 
\end{IEEEproof}

If the system is structurally resilient to DoS attacks and (\ref{DoSToInteg}) holds, the same configuration will automatically make it structurally resilient to SFI attacks. 
However, there might be a cost involved in `turning on' controls to ensure structural controllability, and the system might want to be resilient with the lowest cost. 
This would entail choosing a subset of the columns of $[B_{def}]$, indexed by $\ca{I}$, to maintain structural controllability of $([A_{att}],[B_{def}(\ca{I})])$, while minimizing the cost of the control action. 
It is important to note that structural resilience to DoS attacks guarantees structural resilience only to SFI attacks. 
It does not, in general, ensure structural resilience to arbitrary integrity attacks.

We conclude this section by distinguishing the analysis above with data integrity attacks (for e.g., false-data injection attacks \cite{zhang2020false}). 
We focus on the ability of an adversary to influence the structural representation of the system matrix during state-feedback attacks. 
This corresponds to using structural representations of the matrices $[K_{def}]$ and $[K_{att}]$ to analyze the structural resilience of the systems in Equations (\ref{IntegModel}) and (\ref{SFIntegModel}). 
This approach is independent of the numerical values of the `false-data' and numerical values of the entries of the state-feedback matrices. 
In comparison, assumptions may be needed on magnitudes of error and residual signals, or inverses of certain matrices may have to be computed when characterizing the effect of data-integrity attacks \cite{zhang2020false}. 
Furthermore, if the attack does not influence the system structure, the analysis of structural properties of the system before and after the attack will yield identical results. 
\section{Computational Complexity}\label{complex}

The computational complexity of determining the structural resilience of the system under both DoS and integrity attacks depends on: \emph{i)} determining SCCs in a digraph; and, \emph{ii)} determining a maximum matching in a bipartite graph. 

SCCs in a digraph can be computed using Tarjan's algorithm \cite{tarjan1972depth}, which in the worst-case, is $\ca{O}(|\ca{V}|+|\ca{E}|)$. 
A maximum matching of a bipartite graph can be determined by the Hopcraft-Karp algorithm \cite{hopcroft1971n5}, whose complexity in the worst-case is $\ca{O}(\sqrt{|\ca{V}|}|\ca{E}|)$. 
An extension for determining maximum matchings in more general graphs with the same computational complexity was presented in \cite{micali1980v}. 

\section{Examples} \label{Examples}

In this section, we present multiple examples to illustrate the results in Sections \ref{DoSRes} and \ref{IntegRes}. 
In all the examples, we will assume that $x_1,\dots,x_6 \in \ca{X}_{def}$ and $x_7,\dots,x_{10} \in \ca{X}_{att}$. 
We identify connections between these examples and results that were presented in Sections 4 and 5.
\begin{eg}(\textbf{DoS Attack Resilience}) 
%Consider the digraph of $[A]$ in Fig. (\ref{fig1a}). 
Figure \ref{fig1a} shows the directed graph representation of a system, $\ca{D}([A])$. % with $x_1,\dots,x_6 \in \ca{X}_{def}$ and $x_7,\dots,x_{10} \in \ca{X}_{att}$. % in $\ca{D}([A])$ in Fig. (\ref{fig1a}). 
The SCCs are $(x_1,x_2,x_3), (x_8), (x_4,x_5,x_6,x_7),$ and $(x_9,x_{10})$. 
Inputs need to be asssigned to the first two SCCs, since they are not top linked. 
Every maximum matching of $\ca{B}([A])$ will have $x_8 \in \ca{X}_{att}$ as a right unmatched vertex. 
Thus, the system is not structurally resilient to a DoS attack. 

\begin{figure}
\centering
\subfloat[ \label{fig1a}]{
  \includegraphics[width=0.9 in]{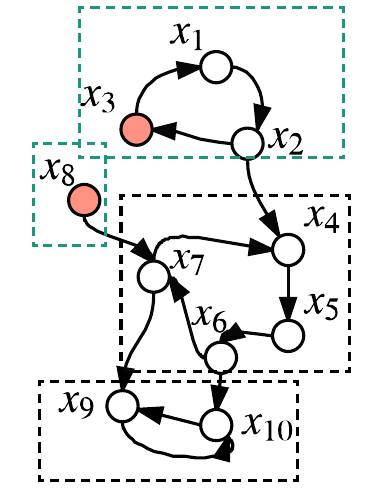} 
} \hfill
\subfloat[ \label{fig1b}]{
  \includegraphics[width=0.9 in]{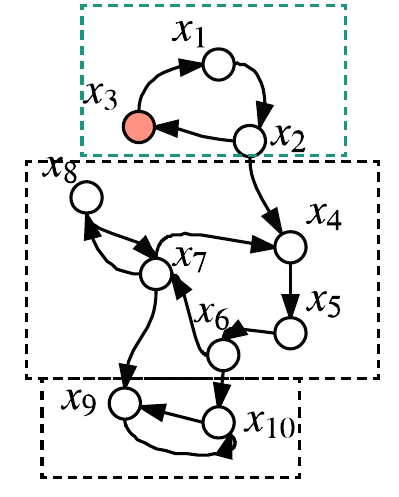} 
} \hfill
\subfloat[ \label{fig1c}]{
  \includegraphics[width=0.9 in]{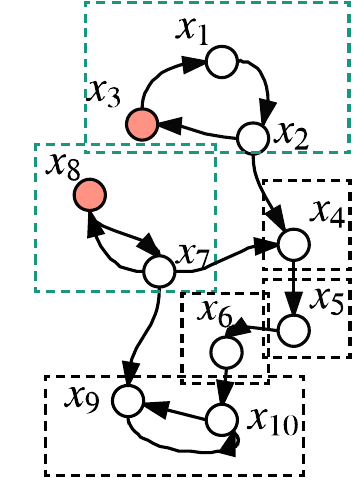} 
} \hfill
\caption{Structural Resilience to DoS Attack} 
\end{figure}

Now, add the edge $x_7 \rightarrow x_8$ to the digraph as shown in Figure (\ref{fig1b}). 
The SCCs are $(x_1,x_2,x_3), (x_4,x_5,x_6,x_7,x_8),$ and $(x_9,x_{10})$. 
Only the first SCC is not top linked, and there is only one right unmatched vertex in every maximum matching, and for some such matching, it is not in $\ca{X}_{att}$. 
Therefore, this system is structurally resilient to a DoS attack. 

If $x_6 \rightarrow x_7$ is removed (Figure (\ref{fig1c})), then $(x_7,x_8)$ is a non-top-linked SCC, which necessitates the assignment of a control to it, making the system vulnerable to a DoS attack. 
%This system is shown in Figure (\ref{fig1c}).
\end{eg}

Example 7.1 illustrates three different structural representations, and SCCs for each representation are shown in Fig. 2. 
In Fig. (\ref{fig1a}), $x_8$ being a right unmatched vertex in all maximum matchings of $\ca{B}([A])$ violates the first condition in Theorem \ref{DoSThmMain}. In this case, $x_8$ is the only vertex in a non-top linked SCC (green dotted box), which also violates the second condition of Theorem \ref{DoSThmMain}. 
The structure in Fig. (\ref{fig1b}) satisfies both conditions of Theorem \ref{DoSThmMain}, and the system can be controlled by connecting the vertex $x_3 \in \mathcal{X}_{def}$ to an input, making this configuration structurally resilient to a DoS attack. 
The structure in Fig. (\ref{fig1c}) violates the second condition of Theorem \ref{DoSThmMain}, since $x_7, x_8 \in \ca{X}_{att}$ form a non-top linked SCC.
\begin{eg} (\textbf{SFI Attack Resilience}) 
In Figure (\ref{fig1a}), if a state feedback adds an edge $x_7 \rightarrow x_8$ %or $x_8 \rightarrow x_9$
, then there is a maximum matching of $\ca{B}([A_{att}])$ with no right unmatched vertices or non-top-linked SCCs in $\ca{X}_{att}$, ensuring structural resilience to a state feedback attack. 
In Figure (\ref{fig1b}), any state feedback $[K_{att}]x$ will add edges to the set $\{x_7,x_8,x_9,x_{10}\}$. 
We know that this graph does not have right unmatched vertices in $\ca{X}_{att}$. 
This ensures structural resilience with the same $[B_{def}]$ as in the DoS case.
\end{eg}

In Example 7.2, the addition of the edge $x_7 \rightarrow x_8$ by a state feedback $[K_{att}]$ to the structural representation in Fig. (\ref{fig1a}) will yield the representation of Fig. (\ref{fig1b}). 
From Example 7.1, we know that the latter representation is resilient to a DoS attack. 
Moreover, since the condition in Eqn. (\ref{DoSToInteg}) will hold, from Theorem \ref{ThmDoSToInteg}, the same $[B_{def}]$ used to ensure structural resilience to a DoS attack will also guarantee resilience to the SFI attack. 

\begin{eg} (\textbf{Integrity Attack Resilience}) 
For $[A_{def}]$ in Figures (\ref{fig1a}, \ref{fig1b}, \ref{fig1c}),  %Figures of DoS
there is a non-top-linked SCC with vertices only in $\ca{X}_{def}$.   
Since controls in $u_{att}$ cannot be assigned to vertices in $\ca{X}_{def}$, the systems are structurally resilient to an integrity attack. 
This conclusion is a consequence of Theorem \ref{ThmInteg}.
\begin{figure}
 \centering
  \includegraphics[width=1.2 in]{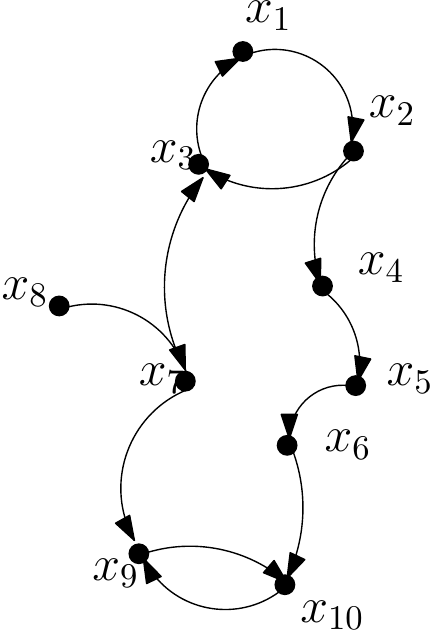} 
\caption{Structural Resilience to Integrity Attack} \label{fig2} 
\end{figure}

For the $[A_{def}]$ shown in Figure (\ref{fig2}), all maximum matchings will have $x_8$ as a right unmatched vertex, and $x_8 \in \ca{X}_{att}$ will be a non-top-linked SCC. 
Moreover, this will be the only non-top linked SCC, which allows us to apply Theorem \ref{CorInteg}.  
Complete controllability by the attacker will be possible by supplying an input to $x_8$, and the system will therefore not be resilient to an integrity attack. 
\end{eg}
\section{Extension to Switched Systems}\label{Switched}

This section introduces a characterization for the resilience of switched linear systems (SLSs) to DoS attacks. 
A switched system comprises a family of subsystems and a rule which governs transitions among these subsystems. 
Each subsystem of an SLS is modeled as a linear dynamical system. 
We use the structured systems framework to obtain a graph-theoretic representation of an SLS. 
In order to gain insight into the operation of the SLS in its constituent subsystems, we construct a \emph{union graph}.   
We show that our results in Section \ref{DoSRes} can be adapted to establish conditions for a structured SLS to be resilient to DoS attacks.  

\subsection{Switched Linear Systems}
A switched system comprises a family of subsystems and a rule that governs switching among them. 
In this paper, we will assume that each subsystem is a linear system, given by: 
\begin{align}
\dot {x}(t)&=A_{\sigma(t)}x(t)+B_{\sigma(t)}u(t) \label{StateEqnSw} 
\end{align}
where $x(t) \in \mathbb{R}^n$ and $u(t) \in \mathbb{R}^p$. 
$\sigma : [0, \infty) \rightarrow \mathbb{M}:=\{1,\dots,z\}$ is a switching signal. 
$\mathbb{M}$ are the \emph{modes} of the system, and $\sigma(t)=i$ implies that the $i^{th}$ subsystem is active at time $t$. 
We make the following assumptions in the sequel. 

\begin{ass}
\emph{i)} The switching signal $\sigma(t)$ does not depend on initial states and controls; 
\emph{ii)} There is only a finite number of changes of mode in every finite time interval; 
\emph{iii)} All pairs of mode transitions are allowed, and there is no constraint on the time the system must spend in each mode. 
\end{ass}

Assumption $8.1.i)$ is standard in the switched systems literature; $8.1.ii)$ is needed to rule out the Zeno phenomenon. 

Let $[A_k]$ and $[B_k]$, $k \in \{1,\dots,z\}$ correspond to the structural realization of matrices $A_k$ and $B_k$ respectively. 
Therefore, $[A_k] \in \{0, *\}^{n \times n}$ and $[B_k] \in \{0,*\}^{n \times p}$. 
We can associate a directed graph to each mode of the system. 
Let $\ca{D}_k=(\ca{V}_k, \ca{E}_k)$, where $\ca{V}_k=\ca{U}_k \cup \ca{X}_k$ and $\ca{E}_k=\ca{E}_{A_k} \cup \ca{E}_{B_k}$, 
% \cup \ca{E}_C \cup \ca{E}_D$
where $\ca{E}_{A_k}=\{(x_j,x_i)| [A_k]_{ij} \neq 0\}$, $\ca{E}_{B_k}=\{(u_j,x_i)| [B_k]_{ij} \neq 0\}$, $k=\{1,\dots,z\}$. 
We now define the notion of a \emph{union graph}.

\begin{df}
The \emph{union graph} of a collection of digraphs $\ca{D}_k:=\ca{D}([A_k],[B_k])=(\ca{V}_k, \ca{E}_k), k=\{1,\dots,z\}$ is given by $\ca{D}:=(\ca{V}_1 \cup \dots \cup \ca{V}_z, \ca{E}_1 \cup \dots \cup \ca{E}_z)$.  
\end{df}
\begin{rem}
Structurally, an edge $e_{ij}$ in the union graph corresponds to a non zero entry in the $(j,i)$ position in at least one of the $[A_k]$ (or $[B_k]$) matrices. 
The absence of an edge $e_{ij}$ from vertex $i$ to vertex $j$ indicates that the $(j,i)$ entry in each of the $[A_k]$ and $[B_k]$ matrices is zero.  
Equivalently, the union graph is a representation of the structured system defined by $([A_1]+\dots + [A_z], [B_1]+ \dots + [B_z])$. 
\end{rem}

We will denote the union graph of structured matrices $[M_1]$ and $[M_2]$ by $\ca{D}([M_1] + [M_2])$, and $[$ $[M_1],[M_2]$ $]$ will denote the concatenation of the matrices $[M_1]$ and $[M_2]$. 
%
%The following is a necessary and sufficient condition for structural controllability of the switched linear system. 
\begin{thm} \cite{pequito2017structural} \label{SwStructCont}
A switched linear continuous time system is structurally controllable if and only if:% the following two conditions hold: 
\begin{enumerate}
\item there exists an edge from an input in the digraph $\ca{D}([A_1] + \dots + [A_z], [B_1] + \dots + [B_z])$ to a state vertex in every non top linked SCC of $\ca{D}([A_1] + \dots + [A_z])$.
\item the bipartite graph $\ca{B}([$ $[A_1],\dots,[A_z],[B_1],\dots,[B_z]$ $])$ has a maximum matching of size $n$.  
\end{enumerate} 
\end{thm}

The authors of \cite{pequito2017structural} showed that if a switching signal ensures structural controllability of the SLS, the property is invariant to the order in which mode transitions occur. 
Therefore, if certain mode transitions are forbidden, then the switching signal can be chosen to satisfy these constraints. 
\begin{eg}
Let an SLS have modes $M_1,M_2,M_3,M_4$, and the transitions $M_2 \rightarrow M_3$ and $M_1\rightarrow M_4$ be forbidden. 
Then, if a switching signal $M_1M_2M_3$ ensures structural controllability, the SLS will be controllable for all mode transitions not involving $M_4$ and not involving $M_2 \rightarrow M_3$ and $M_1\rightarrow M_4$. 
An example of a switching signal that will ensure structural controllability of the SLS is $M_3 \rightarrow M_1 \rightarrow M_2 \rightarrow M_1 \rightarrow M_3$. 
Another example is $M_1 \rightarrow M_3 \rightarrow M_2$. 
\end{eg}

\subsection{Structural Resilience}

We use the union graph representation introduced above to provide a characterization of the resilience of the SLS to denial-of-service attacks.  

%The input in Equation (\ref{StateEqnSw}) is partitioned as
We write $u=\bigl(\begin{matrix}
u_{1}&\dots&u_d&u_{d+1}&\dots&u_{p}
\end{matrix} \bigr)^T$ for the input in Equation (\ref{StateEqnSw}), and 
use $u_{def} \in \bb{R}^d$ and $u_{att}\in \bb{R}^a$ (with $a:=p-d$) to denote $\bigl(\begin{matrix} u_{1}&\dots&u_d\end{matrix} \bigr)^T$ and $\bigl(\begin{matrix} u_{d+1}&\dots&u_{p} \end{matrix} \bigr)^T$ respectively. 
The structural equivalent of Equation (\ref{StateEqnSw}) is:% can be written as:
\begin{align}
\dot{x}(t)&=[A_{\sigma(t)}]x(t)+[B_{\sigma(t)_{def}}]u_{def}(t)+[B_{\sigma(t)_{att}}]u_{att}(t) \label{StructSw}
\end{align}

If $\ca{X}_{def}$ ($\ca{X}_{att}$) denotes the disjoint sets of state vertices that are accessible to the defender (attacker) inputs, then $[B_{k_{def}}]$ ($[B_{k_{att}}]$) will have fixed zeros in rows corresponding to $\ca{X}_{att}$ ($\ca{X}_{def}$).% and $[B_{k_{att}}]$ will have fixed zeros in rows corresponding to $\ca{X}_{def}$. 
During a DoS attack, the inputs in $u_{att}$ are set to zero. 
Structurally, this corresponds to setting every entry of $[B_{k_{att}}]$ to zero for every mode $k$. 

\begin{ass}
The state vertices that the defender and attacker have accesss to remains the same irrespective of the mode of the system. 
That is, the column indices corresponding to $[B_{k_{att}}]$ is the same for every mode. 
\end{ass}
This is a reasonable assumption since an attacker may not have the ability to influence different states of the system at a time-scale faster than that of the switching among modes of the system. 

We formally state the problem that we wish to solve: 
\begin{prob}
Given that the system in Equation (\ref{StructSw}) is structurally controllable before an attack, characterize its structural resilience to a denial of service attack.
\end{prob}

Let $m_{def}$ ($m_{att}$) be the number of right unmatched vertices in $\ca{B}([$ $[A_1],\dots,[A_z]$ $])$ corresponding to $\ca{X}_{def}$ ($\ca{X}_{att}$). 
We are now ready to state the main result of this section. 

\begin{thm}
The switched system is structurally resilient to a denial of service attack if and only if $d \geq m_{def}$ and:
\begin{enumerate}
\item $\ca{D}([A_1]+\dots+[A_z])$ has no non-top linked SCC comprised exclusively of vertices from $\ca{X}_{att}$. 
\item there exists a maximum matching of $\ca{B}([[A_1],\dots,[A_z]])$, denoted $M$, containing every vertex in $\ca{X}_{att}$. That is, $m_{att}=0$ for some maximum matching. 
\item every right unmatched vertex of $\ca{B}([$ $[A_1],\dots,[A_z]$ $])$ in $M$ is connected to a unique input in $u_{def}$. 
\item every non-top linked SCC of $\ca{D}([A_1]+\dots+[A_z])$ contains a vertex in $\ca{X}_{def}$ that is connected to some input in $u_{def}$. 
\end{enumerate}
\end{thm}

\begin{IEEEproof}
If $d <  m_{def}$, then there is some vertex in $\ca{X}_{def}$ that does not have a `dedicated input' needed to ensure structural controllability (Theorem \ref{MinInp}). 

Now consider the case when $d \geq m_{def}$, but $\ca{D}([A_1]+\dots+[A_z])$ contains a non-top linked SCC comprised exclusively of vertices from $\ca{X}_{att}$ or if every maximum matching of $\ca{B}([$ $[A_1],\dots,[A_z]$ $])$ contains some vertex in $\ca{X}_{att}$. 
This would mean that vertices in $\ca{X}_{att}$ would have to be connected to a control in $u_{def}$, which violates our assumption that controls in $u_{def}$ can only be connected to states in $\ca{X}_{def}$. 
The last two conditions are needed to ensure structural controllability of $([A_1],[B_{1_{def}}],\dots,[A_z],[B_{z_{def}}])$. 
Therefore, if any of the conditions are violated, the system will not be structurally resilient to a DoS attack. 
This proves necessity. 

For sufficiency, it is clear that if all the conditions are met, there exists a control configuration which ensures structural controllability even when the system (defender) can control only a subset of the states (i.e., those in $\ca{X}_{def}$), and other states (i.e., those in $\ca{X}_{att}$) cannot be directly accessed. 
\end{IEEEproof}

The above result presented a characterization of the resilience to DoS attacks by providing necessary and sufficient conditions in terms of unmatched vertices of bipartite graphs and strongly connected components of directed graphs that represented the switched system. 
Furthermore, this result is independent of the order of switching among modes of the system, and the time spent in each mode.  

\section{Conclusion and Future Work} \label{Future}
This paper studied the structural resilience of CPSs to DoS and integrity attacks using linear structured systems and graph theory. 
Conditions for the system to be resilient were characterized in terms of unmatched vertices of bipartite graph and connected components of directed graph representations of the structured system. 
An extension to the linear structured switched systems case was studied and conditions needed to establish the resilience to denial of service attacks were presented. 
These conditions were independent of the order of switching among modes and the time spent in each mode.

%Throughout this paper, we have assumed that the system and the attacker have access to disjoint sets of nodes. % that are disjoint. 
One direction of future research is to study structural resilience when there is a set of nodes accessible to both, defender and attacker. 
Another topic of interest is to study the design of `repair mechanisms' so that a defender may be able to add or remove edges to a directed graph representation of the system ensure resilience to an adversary in an adaptive manner. The sets of states accessible to the defender and attacker in this scenario may be time-varying. 
For switched systems, future work will study the case when the sets of states accessible to the defender and attacker is different for each mode. 
%Further, there were no restrictions on the allowed mode transitions or on the duration of time the system could spend in each mode. 
Extending our work to incorporate restrictions on the allowed mode transitions or on the duration of time the system could spend in each mode is another area of interest. 
Alternatively, one could associate probabilities with the transitions from one mode to another, and use this to develop a notion of probabilistic structural resilience for switched systems. 

\bibliographystyle{IEEEtran}
\bibliography{ThesisBib}
\end{document}